# On Geo Location Services for Telecom Operators

DMITRY NAMIOT[1], MANFRED SNEPS-SNEPPE[2]
[1]Lomonsow Moscow State University, Moscow, Russia
dnamiot@gmail.com
[2]Ventspils University College, Ventspils, Latvia
manfreds.sneps@gmail.com

*Abstract:* - This paper presents location based service for telecom providers. Most of the location-based services in the mobile networks are introduced and deployed by Internet companies. It leaves for telecom just the role of the data channel. Telecom providers should use their competitive advantages and offer own solutions. In this paper, we discuss the sharing location information via geo messages. Geo messages let mobile users share location information as signatures to the standard messages (e.g., email, SMS). Rather than let some service constantly monitor (poll) the user's location (as the most standalone services do) or share location info within any social circle (social network check-in, etc.) The Geo Messages approach lets users share location data on the peer to peer basis. Users can share own location info with any existing messaging systems. And messaging (e.g., SMS) is the traditional service for telecom.

## 1 Introduction

It is obvious, that the question "where are you" is one of the most often asked during the communications. 600 billion text messages per year in the US ask "where are you?" – as per Location Business Summit 2010 data. A vast amount of mobile services is actually being built around this question so their main feature is a user's location exchange [1]. In the most cases, the location exchange presents the ability to the mobile user (mobile phone owner) to save own location information in the some special place (e.g., special data store, supported via some mobile application). So, the second party in this exchange process will be able to read saved data. But it means, of course, that a user must be registered with this service and (in the most cases) download a priori some special application. What is even more important here – both parties in the location exchange process must use the same service too. In practice, this leads to a parallel coexistence a set of conceptually similar services.

There are several models for location information sharing in the mobile services. On the first hand, it is passive location monitoring. The typical example is Google Latitude [2]. The word "passive" here describes the system from the end-user's point of view. Passive location sharing model does not require specific actions from mobile users. Accumulated data become available some API. The privacy is probably the biggest issue with this approach. All potential users should be aware that some third party tool is constantly monitoring their location and saves it on the some external server. And, of course, the shortened life of the handset's battery is the second biggest issue with this approach.

Note that in many cases this is not necessarily associated with the installation the special application. Such monitoring can be done and the service provider (mobile operator, etc.). But because we are talking about data exchange, there is a big question: how to use such automatically collected data in third party services? Actually, we suggest a possible model for this use case below.

Another model is a voluntary location sharing. The typical example is check-in [3]. Check-in is a special type for the record (status) in some social network. It could be an active (e.g. Foursquare), when the user directly sets his/her current location or passive (e.g., Twitter), when location information could be added as an additional attribute to the current status. Of course, for sharing location information both parties must be registered in the same network. And here we can see "all or nothing" problem with location sharing. Shared location info could be visible to all friends, but in the real cases for most of them it is just a noise. The location info could be actually interested only for the physical friends. E.g., for the majority of twitter followers my location info (Foursquare status in Twitter time line), is just a noise.

The idea of the signed geo messages service (geo mail, geo SMS) is based on the ability to add user's location info to the standard messages like SMS or email [1]. So, as the answer for the above-mentioned question 'Where are you?', someone may just send a message. And the target party does not need to use any additional service in order to get information about the sender's location. He will simply read SMS or email.

Speaking more broadly, this service separates location information and identity information. The message itself contains the identity. And location sharing data contain the location information only. Only the combination of both elements lets us associate location data with identity.

It is obviously peer to peer sharing and it does not require any social network. For example, the geo signature may have a form of the map with the marker at the shared location. And what is important here – the map itself has no information about the sender and recipient. That information exists only in the message itself. The map (marker) has no information about the creator for example. That is all about the privacy.

This model is probably the easiest way for sharing location information. It does not require any application downloading or registration in social networks from the potential users. This approach provides a smooth extension for the existing communication services.

There are several services that implement Geo Messages approach. Originally, they were described in [1]. And this paper summarizes the latest development, as well as discusses the possible extensions.

## 2 Problem Formulation

The main idea behind Geo Messages is how to deliver location info via the standard messaging (SMS and Email). This approach borrowed the ideas from SMS based content delivery. Typically, when the mobile users request some service via SMS it means users are actually getting as the response some link within the text message. This link leads to the downloading service for pictures, ringtones, etc. And this approach uses the simple fact that all native SMS clients nowadays are smart enough to discover links (just http://something_is_here text chunks) within the text and allow one click Internet access for opening that link. So, for delivering location information we can use the same approach.

The location info could be presented as a link, leads to the appropriate map. So, as soon as the sender will be able to automatically add such a link to the message, the receiver will be able with just one click open the map with the sender's location. By default, this map will show two POI (point of interests) - the sender's location and the receiver's location. Alternatively, we can provide a link to some specially created landing page, probably with a map or any other location related info.

Our original development targeted feature phones and was implemented as an application for SIM-cards (Java-card applet). It includes the following steps:

1. The location info could be requested right from the SIM-card (smart card) as Cell ID info. This information exists always and Java-card applet can read it (local info).

2. Cell ID information could be translated into "human"-readable form of (latitude, longitude) pair. There are several public services over the Net that let us do that. The typical example is OpenCellID [4]. Actually, it is just a public HTTP based API.

3. Using the data obtained in the step 2, we can create a link to the map. Original development used Google Static Map. The Google Static Maps API lets applications embed a Google Maps image on your webpage without requiring JavaScript or any dynamic page loading. In our case, Google Static Maps API lets us build a map (actually – an image with a) based on the latitude – longitude pair obtained through the step 2. For the smartphones, we can create the similar link with Google Maps API (there are no more JavaScript's limitations).

4. URL shortening service could be deployed. In order to make sure our geo-related URL's complies with SMS restrictions (simply – they are no more than 140 symbols) we can deploy URL shortening service and make our signature smaller. It is very important also, that the URL shortening service lets obtain statistics for the deployed URLs. In case of mobile messaging with geo-coding it leads to the context-aware statistics (when and where some link has been opened)

5. In order to add our location aware URL to the message (to SMS or to email) we will deploy URI Scheme for GSM Short Message Service and The mailto URL scheme [1]. So, our final step included dynamical generation of the mobile web page with links for messaging:

sms:?body=our_geo_aware_URL
and
mailto:?body=our_geo_aware_URL

As soon as the mobile user will hit one of the links, the native (it is very important!) messaging client (e.g., the native SMS client) will be launched with the text (body) field being pre-populated with the given URL. So, it is enough just to select the target phone (address) from the address book, add some text (optionally, of course) and send the message. After all, this service presented a mobile mashup (mobile web mashup) that passes user through the series of screens where the last one offers for the user customized messages sending links. And the whole process is

a) completely automated
b) does not require any authorization in external services
c) completely portable and will work on any mobile phone

For HTML5 applications, we can use its geo-capabilities [6]. The modified web client is illustrated in Figure 1.

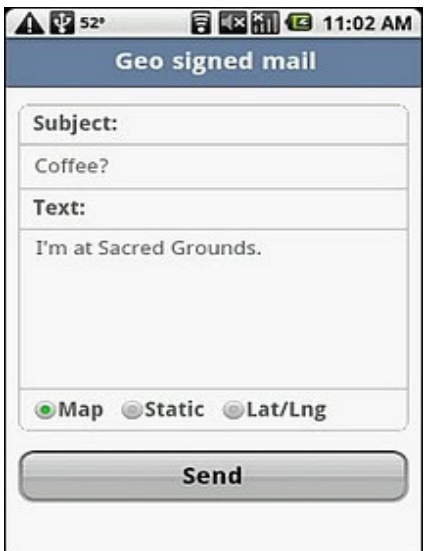

Fig. 1 Geo Mail client

The link in the signature may include a map with some pin, map's snapshot (static picture with map's snapshot), a special landing page or just a text with geo-coordinates. The text is useful for putting data into navigation devices. The landing page is, obviously, a direct way for telecom operator to monetize this location service. The landing page could be generated automatically and present some mini-portal for geo-point in question. It may include a map, some text information and advertising. Figure 2 illustrates the delivered map.

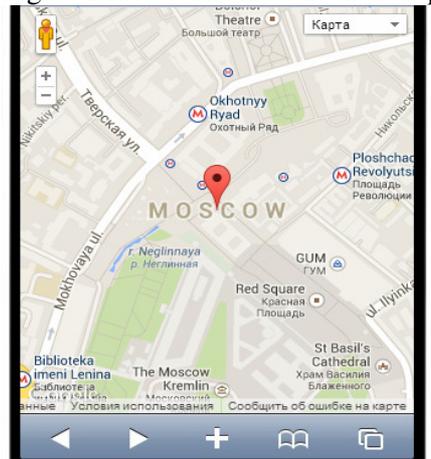

Fig. 2 The delivered map

Obviously, it is a user-friendly sharing (the location is visible as a part of the message). But by the same principles, we can deliver location information in the header of the message. We will explain below how this feature could be implemented.

And the key question is how to get location information for this modified messaging client. Originally [1][5], we present our solutions in the web-mashup style. Below we will try to show that this solution could be an ideal application for telecom operators. It can let them replace Over-The-Top messages with the classical telecom services and avoid the perspective to be a dump pipe for Internet companies.

## 3 Geo Messaging for Telecom

The main idea behind geo-messages is actually very simple. Shortly, geo message means pre-filled (pre-populated) messaging client. The key question is how to obtain data for this initial filling? And here we see the important role of OMA SUPL [6]. SUPL defines a way, the mobile device can obtain location information via the own mobile network.

In GSM networks, we can mention two different types of protocols and localization measurements: Control Plane (C-plane) and User Plane (U-plane). The main difference is in the underlying networks.

C-Plane protocols work in the signaling layer. It means that we do not need a special support form the end-user device point of view. For example, the device does not need TCP/IP support. C-Plane does not touch the application layer. By this reason, this approach is actually very fast.

A disadvantage of C-Plane localization is the need for network-specific upgrades on the provider's site. Also, a provider needs to reserve dedicated channels and frequencies for positioning, depending on the measurement used for locating a handset [7]. The main usage (deployment) for C-Plane protocols is, for example, the E-911. It is obvious, because not every mobile terminal supports TCP/IP but it still needs the localization.

Vice versa, U-Plane protocols work in the application layer. They are independent of the underlying network type and use TCP/IP for positioning determination. For example, they will work in Wi-Fi networks too. SUPL (Secure User Plane Location) is U-Plane protocol. SUPL works is Wi-Fi and GSM networks. As per its definition, SUPL is so-called "Enabler", which uses standards and protocols "where available and possible" [8] to determine the position of a mobile device. SUPL was developed by the OMA (Open Mobile Alliance) to support the creation of interoperable end-to-end mobile services to standardize the communication between the SUPL network and a client device. So, originally, it was suggested as a portable solution for application development. SUPL network here is operator's network (e.g., GSM mobile network). The client device is some mobile terminal (mobile phone). It works in the U-plane. SUPL support various positioning methods: Assisted GPS, Autonomous GPS, E-OTD, Enhanced Cell/Sector ID [9].

There are two modes for using SUPL depends on the initiator. The location request could be initiated either by the mobile terminal (SET in SUPL's terminology), or by the network itself.

In the latest version SUPL introduces triggered positioning procedures. In other words, it is possible to set actions when a mobile terminal entered an area, or send periodic signals of the SUPL enabled device's location. It is illustrated in Figure 3.

There are four different ways for creating (initializing) SUPL session: OMA Push, SMS, UDP/IP, SIP Push. Actually, developers can choose any of the above listed approaches depending on the mobile terminal capabilities and network's circumstances. For our tasks we are interested in UDP/IP sessions.

To create a terminal-initiated session, a mobile terminal (an application) initializes a TCP/IP connection to Secure Location Platform (SLP). To create a network-initiated SUPL session to a mobile the SLP can send a push message to the device with the IP address of the SLP. The mobile device then has to initialize a TCP/IP connection to the SLP with the provided IP address.

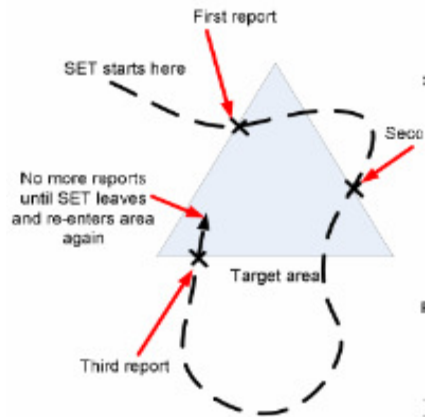

Fig. 3 SUPL triggers [8]

The sequence diagram for the terminal-initiated sessions is presented in Figure 4.

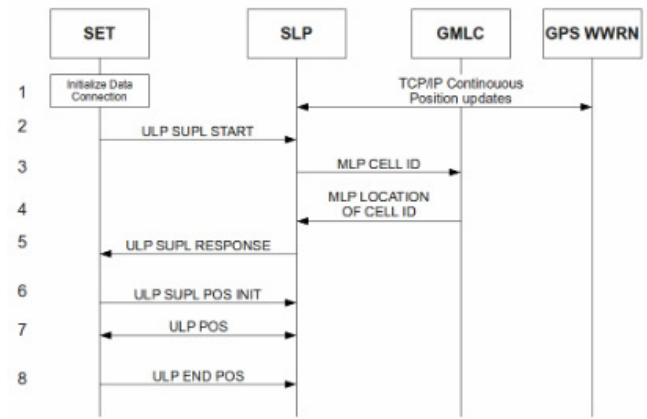

Fig. 4 SUPL UDP/IP session [10]

By our opinion, SUPL being initially oriented to mobile developers is actually a first class instrument for telecom operators. They should use it with the redesigned messaging client. SUPL network is under full operator's control. So, it looks pretty logical to add own client for this network. This client is just an extension of the existing SMS client, for example. This geo-SMS client will open pre-filled SMS messenger. And default text could be the above-mentioned geo-signature exactly. This client will initiate SUPL session, obtain geo coordinates, convert then into one of the above mentioned forms. Note, that this new client could easily use a bit more optimized strategy than simply initiate a new session with SLP during the each call. We can use the accelerometer for detecting device movement and use previously requested data as soon as the device is not moved (or does not move significantly). It is a pretty standard task for smartphones – implement some form of inertial navigation system [11]. And accelerometer here is a good fit due to its low energy consumption [12].

The above mentioned paper [5] extends a generic one to one relation in geo messages with a one-to-many model. Any user can simultaneously support several peer-to-peer location sharing flows. But because the basic principles are the same, this new model could benefit from SUPL too.

In the description above we've followed to the original development of geo-messages. It was a human readable link in the text message. But we can send geo-location in the header for the email too. It could be processed programmatically. This approach could be useful for some automated tasks. For example, an automated email could be used as a replacement for HTTP post with data, etc. There is a standard way of adding new headers for email and HTTP requests. It is so-called "X-" convention [13]. We can follow to GeoRSS concept [14] and directly place a pair of (latitude, longitude) as a new custom header. E.g.:

X-GEO: latitude longitude

Alternatively, for setting a special header we can follow to the concepts of geo hash [15] and geo-cookies [16].

The Geo-hash is a simple method for geo-coding a pair of latitude/longitude coordinates into a shorter hash. This encoding is achieved by recursively dividing the latitude and longitude into two intervals. In the first step intervals are -900 - 00 and 00 - 900. The lover interval corresponds to the binary 0 and the upper interval corresponds to binary 1. In the second step the each interval should be divided again. E.g., it could be 00 to 450 and 450 to 900, and so on. The two resulting bit sequences are then alternately interleaved. The result could be encoded and presented as a string. Two hashes with the same prefix present the same region. So, we can directly use geo-hash for proximity estimation. Also, this approach lets easily change the precision for geo-coding. It could be used in the traditional privacy-related settings for location based systems [17]. E.g., it could be an area-wide geo-coding, city-wide geo-coining, street level, etc. So, depends on the privacy (security) settings, the users may see the same shared location with the different precision [18]. It is so-called location obfuscation [19].

By our opinion, the modified messaging client will be able to replace OTT messengers [20].

We can propose a more interesting model with dynamic messages. In this case, our message will have SET identification instead of the static location. The idea allows for messaging client (e.g., SMS client) to request the current location any time the message is opened. Think, for example, about this use case. User A is going to meet user B. User A sets an initial messaging about his intention to come. This message (e.g., SMS) contains SET ID for user A. User B opens the message and can see the current location for user A. So, we can display on the map the current locations for both users A and B, estimate the distance, estimate the time for approaching, etc. And it is not an application. Vice versa, it is a "standard" messaging client.

OMA introduced support for indoor navigation in its recent Enablers Secure User Plane Location (SUPL 3.0) and LTE Positioning Protocol Extensions (LPPe 1.0) [21].

The goals of SUPL 3.0 and LPPe 1.0 are to improve the user experience through better service and new features, specifically including, improved Indoor Location Accuracy. As an example for the special requirements arising from indoor location issues we can mention the support for floor level information as well as the use of relative instead of global coordinates.

SUPL 3.0 describes the following blocks for indoor navigation support:

1) Decentralized Location Server (D-SLP: Discovered SUPL Location Platform) for Assistance Data Delivery and Position Calculation.
2) Positioning Protocol supports indoor navigation assistance data (map information, etc.).

D-SLP is an additional element of H-SLP (Home SUPL Location Platform). The idea is to introduce a special server for Indoor Positioning support. So, SET (mobile terminal) may choose a special source for indoor data. Also, it lets vendors add own D-SLP services for own venues. The common schema for access follows to the following algorithm:

1) The SET discovers a local SLP (D-SLP) which is able to provide Indoor Positioning service within a defined service area (e.g., within a shopping mall).
2) The SET requests authorization for accessing the D-SLP from its home server (H-SLP).
3) The H-SLP authorizes access within a defined service area, access network, and time window.
4) While the SET is within the service area, time window and the authorized access network of the D-SLP, it may access the D-SLP and obtain Indoor Positioning Services.

For our explanation, it is important that for the indoor environment, both parties in the message exchange will be covered by the same local SLP. It means, we can continue to share location info (indoor location info in this case) with the messaging.

Another important step is the possible replacement for so-called server-side push with SMS for indoor proximity notifications.

Server-side push (or cloud messages) is a service from mobile OS vendors for sending notifications to mobile users. For example, Google Cloud Messaging for Android (GCM) is a service that allows developers to send data the own server to users' Android-powered device. This could be a lightweight message telling your app there is new data to be fetched from the server (for instance, a movie uploaded by a friend), or it could be a message containing up to 4kb of payload data (so apps like instant messaging can consume the message directly) [22].

The GCM service handles all aspects of queuing of messages and delivery to the target Android application running on the target device. GCM is completely free no matter how big your messaging needs are, and there are no quotas. But of course, it is free from the vendor's point of view. For telecom this OTT message has got some cost, of course.

There are conceptually similar services from other vendors (e.g., Apple, Microsoft, Nokia). Architectures of these push notification services have common features. On the first hand, application servers send a notification message with an intended receiver (or the target mobile device) to one of the cloud-based messaging servers. Messaging servers push the message to the target mobile device. The push notification service eliminates the needs of application servers to keep track of the state of a mobile device (i.e., online or offline). Furthermore, mobile devices do not need to periodically probe (poll) the application servers for messages. It reduces the workloads of the application servers and seriously accelerates the mobile application development. Conceptually, any such service replaces telecom notifications (e.g., SMS).

For Bluetooth tag the distance estimation could be based on the ratio of the tag's signal strength (RSSI) over the calibrated transmitter power. The power is the known measured signal strength in RSSI at 1 meter away. Each tag (e.g., iBeacon in case of iOS) must be calibrated with this power value to allow the accurate distance estimation. The iBeacon output power is measured (calibrated) at a distance of 1 meter. Let's suppose that this is $R_1$. The listening device will measure the RSSI of the device. Let's suppose it is $R_2$. Since these numbers are in dBm, the ratio of the power is actually the difference in dB. So:

$$dBm\_ratio = R_1 - R_2 \quad (1)$$

To convert that into a linear ratio, we use the standard formula:

$$LinearRatio = 10^{\wedge}(dBm\_ratio / 10) \quad (2)$$

If we take into account the conservation of energy, then the signal strength must fall off as $1/r^2$ (r here is a distance). So:

$$R = R1 / r^2 \quad (3)$$

$$r = \sqrt{LinearRatio} \quad (4)$$

For indoor location based services push notifications are very often used as a core mechanism for proximity-based information delivery. For example, Bluetooth Low Energy tags from Apple (iBeacons) could be used by mobile applications for proximity detection (Figure 5).

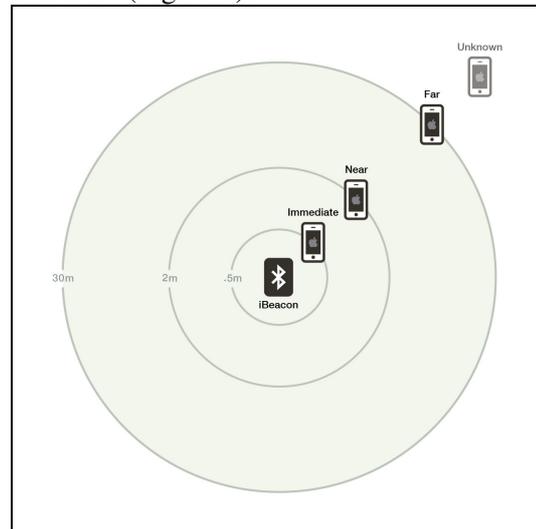

Fig. 5 RSSI-based distance

As soon as proximity is detected, the service will send push notification to mobile user (actually – to some application installed on the user's device). As we wrote above is vendor-based notification (yet another OTT service). With D-SUPL we can use it as a proximity sensor and send notifications (e.g., SMS rather than OTT messages) to all mobile users covered currently by the current D-SUPL. And any such message could have location related data as a signature again. It is yet another example where the telecom operator can replace OTT messages with own stuff (SMS in this case). For OTT messages subject of subscription is some mobile application (in other words, mobile users should previously install an application), for SMS messages subject of subscription is the mobile device itself. We do not need an application. And having location information in the signature lets us provide a consistent experience for our users – there is a constant place for location data for both outdoor and indoor applications.

We should mention in this context also other proposed standards for location measurements. One notable example is GEOPRIV [24]. RFC 3963 describes the basic architecture for GEOPRIV. But at this moment, GEOPRIV describes in RFC3693 only a top-level concept of what is required for a secure transfer of geographical localization data within a network. It does however not define any protocols or data formats on the implementation level. So, we think that OMA SUPL is the most developed standard at this moment and the most suitable for the telecom operators.

## 4 Conclusion

Originally, OMA Secure User Plane model was proposed for application development. In this paper, we explain SUPL usage for telecom operators. By our opinion, this model could be used as a main tool for telecom operators with the idea to redesign messaging clients. Adding the geo location sharing to messaging could be easily adopted by mobile users and can present a winning application for telecom providers.